\documentclass[slac_one]{revtex4}
\usepackage{graphicx}
\usepackage{fancyhdr}
\def\fig#1{fig.~{\ref{#1}}}

\newcommand{\BlackHat}{{\sc BlackHat}}
\newcommand{\lb}{{\bar l}}
\newcommand{\qb}{{\bar q}}
\newcommand{\eps}{\epsilon}
\newcommand{\ep}{\epsilon}
\newcommand{\e}{\epsilon}

\newcommand{\tree}{{\rm tree}}
\newcommand{\cg}{c_\Gamma}

\begin{document}

MIT--CTP 3967 \hfill  SLAC--PUB--13355 \hfill UCLA/08/TEP/26
 \hfill Saclay-IPhT--T08/123
%
\title{One-Loop Multi-Parton Amplitudes with a Vector Boson for the LHC}

\author{C.~F.~Berger${}^{a}$, Z.~Bern${}^b$\footnote{Presenter at 34th
International Conference on High Energy Physics, Philadelphia, USA,
July 2008.},
L.~J.~Dixon${}^c$, F.~Febres Cordero${}^b$, D.~Forde${}^{b,c}$, H. Ita${}^b$, 
D.~A.~Kosower${}^d$ and D.~Ma\^{\i}tre${}^c$}%
\affiliation{\centerline{${}^a${Center for Theoretical
Physics, Massachusetts Institute of Technology,
      Cambridge, MA 02139, USA}} \\
\centerline{${}^b$Department of Physics and Astronomy, UCLA, Los Angeles, CA
90095-1547, USA} \\
\centerline{${}^c$Stanford Linear Accelerator Center, Stanford University,
             Stanford, CA 94309, USA} \\
\centerline{${}^d$Institut de Physique Th\'eorique, CEA--Saclay,
          F--91191 Gif-sur-Yvette cedex, France}\\
}

\begin{abstract}
$\null$ 
\vskip -1. cm In this talk, we present the first, numerically stable, results
for the one-loop amplitudes needed for computing $W,Z$ + 3 jet cross
sections at the LHC to next-to-leading order in the QCD coupling. We
implemented these processes in \BlackHat{}, an automated program based
on on-shell methods.  These methods scale very well with increasing
numbers of external partons, and are applicable to a wide variety of
problems of phenomenological interest at the LHC.
\end{abstract}

\maketitle
\thispagestyle{fancy}


Particle physicists eagerly await the new physics, beyond the Standard
Model, that is expected to emerge at the LHC.  In many cases, the
Standard Model will produce background events that can obscure the
signals of new physics.  Uncovering and understanding these signals
will make use of detailed kinematic constraints (such as several
identified jets, cuts on missing transverse energy, etc.) and reliable
predictions for background processes.

Leading-order cross sections in QCD suffer from large uncertainties, and
do not suffice for quantitative knowledge of backgrounds.
Next-to-leading order (NLO) corrections lead to a significant
improvement.  As reviewed in ref.~\cite{NLMLesHouches}, there are a
large number of processes of interest involving many final-state jets.
Developing methods for computing
such processes has involved a dedicated effort over many years, also
described in ref.~\cite{NLMLesHouches}.
The main bottleneck to NLO computations with four or more final-state
objects (including jets) has been in evaluating one-loop (virtual)
corrections.  In this writeup we present results from
\BlackHat{}~\cite{BlackHatI}, which is one of a new generation of
programs based on on-shell methods.  Other numerical efforts along
similar lines are described in refs.~\cite{OPP,OtherUnitarityNumerical,
OtherNumericalDDim}.  These methods~\cite{UnitarityMethod,
DDimUnitarity, Zqqgg, BCFUnitarity, Bootstrap, BCFCutConstructible, 
Genhel, OPP, Forde, OnShellReview} offer excellent scaling
properties as the number of external legs increases, promising a
general solution to the problem of calculating one-loop amplitudes
with a large number of final state partons.

\begin{figure}[t]
\centering
\parbox{7 cm}{%
\vskip 34pt
\includegraphics[width=7cm]{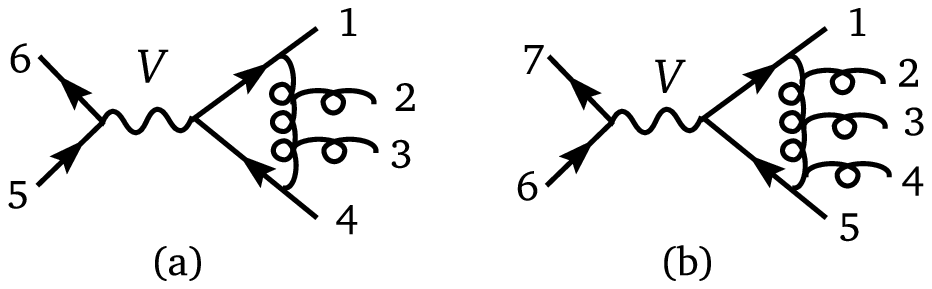}
\caption{Sample leading-color diagrams for the amplitudes 
(a) $q g g \qb \, \lb l$ and (b) $q g g g\qb \, \lb l$.  
The leptons, $\lb,l$ couple to the quarks via a vector boson, $V$. 
Analytic expressions for the six-point 
 amplitudes (a) are given in ref.~\cite{Zqqgg}.}
\label{Z45DiagramsFigure}}
\qquad
\parbox{9 cm}{%
\includegraphics[width=9cm]{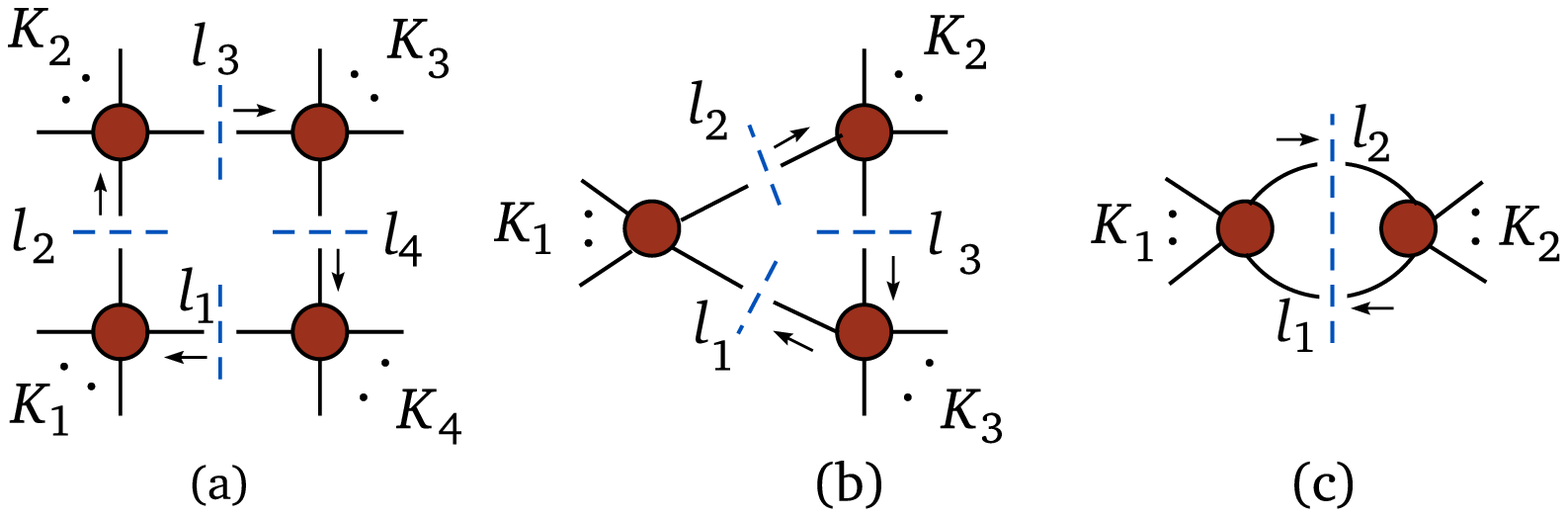}
\caption{The quadruple (a), triple (b) and double (c) cuts used to
evaluate the coefficients of the box, triangle and bubble integral
functions.  The momentum of a cut line satisfies the on-shell 
conditions, $l_j^2 = m_j^2$, where $m_j$ is the mass of the cut line.}
\label{CutsFigure}}
\end{figure}

\BlackHat{} is an automated program generating numerical values for one-loop
amplitudes at given kinematic points.  The on-shell formalism it
uses is distinct from Feynman diagrams.  The latter are inherently
gauge non-invariant, allowing unphysical states to circulate in the
loops.  Large gauge-dependent cancellations between terms make a
conventional diagrammatic approach impractical and numerically
unstable for sufficiently many external states.  (Analytically, the
conventional approach suffers from an explosion in intermediate-stage
computational complexity even with only a modest increase in number of
external legs.)  On-shell methods bypass these cancellations by
building new amplitudes using only simpler on-shell amplitudes as
input. \BlackHat{} has been validated for one-loop six-, seven- and
eight-gluon amplitudes, demonstrating excellent speed and numerical
stability~\cite{BlackHatI}.  In this talk we present results for the
leading-color contributions to one-loop $q ggg \qb \, V$ amplitudes,
where the vector boson $V$ decays into a pair of leptons.  Sample
diagrams for these amplitudes are shown in \fig{Z45DiagramsFigure}(b).
This is a key contribution to NLO corrections to vector boson
($\gamma^*$, $W$, or $Z$) production in association with three jets,
which is an important background to searches for
supersymmetry~\cite{WishList}.  Here we present `primitive'
amplitudes, which can be converted easily to
amplitudes for any choice of $V$, by multiplying by appropriate
propagator and coupling-constant factors~\cite{Zqqgg}.


On-shell methods use the unitarity and factorization properties of
amplitudes in quantum field theories to systematically construct
amplitudes with larger numbers of loops or legs. Terms with branch
cuts can be determined by evaluating unitarity cuts in four
dimensions.  The unitarity method~\cite{UnitarityMethod} provides a
systematic means for constructing these terms directly from tree
amplitudes.  Recent
refinements~\cite{BCFUnitarity,BCFCutConstructible,OPP,Forde},
exploiting complex momenta, greatly enhance the effectiveness of
generalized (multiple) cuts~\cite{Zqqgg}.  By restricting the momenta
of the cut lines to four dimensions we may use powerful spinor
formalisms.  This procedure drops rational terms, which could be
retained by evaluating the cuts in $D$
dimensions~\cite{DDimUnitarity,OtherNumericalDDim}.  Alternatively we can
extract the rational terms using on-shell recursion, developed by
Britto, Cachazo, Feng and Witten at tree level~\cite{BCFW}, and
extended to loop level in refs.~\cite{Bootstrap,Genhel}.

Any dimensionally-regulated one-loop amplitude, with 
four-dimensional external momenta, may be written as,
\begin{equation}
A_n = \sum_i d_i I_4^i + \sum_i c_i I_3^i + \sum_i b_i I_2^i 
+ R_n + {\cal O}(\epsilon)\,.
\label{IntegralBasis}
\end{equation}
\vskip -.2 cm 
\noindent
The scalar integrals $I_{2,3,4}$, respectively bubbles,
triangles, and boxes, are known functions~\cite{IntegralsExplicit}
and contain all the amplitude's branch cuts.
(If massive particles propagate in the loops there are also tadpole
contributions.) The coefficients of these integrals, $b_i, c_i$, and $d_i$, 
are rational functions of the external momenta, which 
may be extracted from generalized unitarity cuts
using four-dimensional values for the loop momenta.
The remaining function, $R_n$, is also rational 
and is computed in \BlackHat{} 
using on-shell recursion.

The box integral coefficients are the most straightforward to compute.
Imposing four on-shell conditions on a one-loop integrand, as shown
in \fig{CutsFigure}(a), while keeping the loop momentum in four
dimensions, freezes the integration completely.
This procedure isolates the coefficient of
a single box integral uniquely, allowing for a simple
evaluation~\cite{BCFUnitarity}.  The coefficient is
given by the product of four tree amplitudes at the corners of the box,
\begin{eqnarray}
d_i &=& {1\over 2} \sum_{\sigma = \pm} 
 A^\tree_{(1)} A^\tree_{(2)} A^\tree_{(3)} A^\tree_{(4)} 
\Bigr|_{l_j = l_j^{(\sigma)}} \,,
\label{QuadCutSolution}
\end{eqnarray}
\vskip -.2 cm 
\noindent
where the cut loop momenta $l_j^{(\pm)}$ are the two solutions to the
quadruple-cut on-shell conditions. 

Triangle coefficients are obtained from the triple cuts shown in
\fig{CutsFigure}(b).  Here, the cut conditions no longer freeze the
integration completely; one degree of freedom is left over. To evaluate
the triangle coefficient we use the analytic parametrization
method of Forde~\cite{Forde}, along with the Ossola, Papadopoulos and
Pittau~\cite{OPP} procedure of subtracting previously-computed box
contributions from the triple cut.  This eliminates unwanted
poles from the complex plane, rendering the computation
of these coefficients more numerically stable.
For the case of bubble coefficients, two
degrees of freedom remain after imposing two cut conditions, as shown
in \fig{CutsFigure}(c).  We refer the reader to
refs.~\cite{OPP,Forde,BlackHatI} for details.  Spinor-based methods have
also been developed for computing triangle and bubble coefficients
analytically~\cite{BCFCutConstructible}. 

At most phase-space points, ordinary double-precision arithmetic
suffices to obtain a relative accuracy of $10^{-5}$ or better, even
for the most complicated of the amplitudes.  This is far better than
is needed in a realistic NLO calculation, given all the other
uncertainties.  Nonetheless, for a small percentage of phase-space
points, numerical instabilities due to roundoff error do arise,
because of large cancellations between different terms in
eq.~(\ref{IntegralBasis}).  Such points typically feature numerically
small values for a Gram determinant.  To identify these points
dynamically we require that all spurious singularities cancel amongst
bubble coefficients, and that the known coefficient of the $1/\e$
singularity generated by bubble integrals be correct.  Whenever the
result at a given point fails these criteria, we recalculate the
amplitude at higher precision.  In practice, this step has only a
modest impact on the overall speed of the program, given the small
fraction of unstable points.


\begin{table}
\caption{\label{VectorBosonTable} Numerical results for leading-color
amplitudes with an intermediate vector boson at the six- and
seven-point momenta given in eqs.~(9.1) and (9.3) of
ref.~\cite{Genhel} (for renormalization scale $\mu=6$ and 7,
respectively). The numerical values for the six-point amplitudes
agree with the analytic results of
ref.~\cite{Zqqgg}.  Our helicity labels follow an all-outgoing
convention.  The three values on each line are the $1/\e^2$,
$1/\e$ and finite contributions to the amplitudes, normalized by the
tree amplitude. 
 }
\begin{center}
\begin{tabular}{||l||c|c|c||}
\hline
\hline
Helicity& $\; 1/\e^2 \;$ & $1/\e$ & $\e^0$ \\
\hline
\hline
   $\; {1_q^+}\, {2_g^+}\, {3_g^+}\, {4_\qb^-}\, {5_{\lb_{\vphantom{A}}}^-}\, 
   {6_{l}^+}  $
&  $\; -3.00000 \; $
&  $\; -5.86428 - i\, 6.28319 \; $ 
&  $\; -7.50750 - i\, 12.90848 \; $ \\
\hline
$\; {1_q^+}\, {2_g^+} \,{3_g^-} \,{4_\qb^-} \,{5_{\lb_{\vphantom{A}}}^-}\, 
   {6_{l}^+}  $
& $\; -3.00000 \; $ 
& $\; - 5.86428 - i\, 6.28319 \; $ 
& $\; \hphantom{-} 0.10598 - i\,11.74860  \; $ \\
\hline
$\; {1_q^+}\, {2_g^-}, {3_g^+} \,{4_\qb^-}\, {5_{\lb_{\vphantom{A}}}^-}\, 
   {6_{l}^+}  $
& $\;  -3.00000\; $ 
& $\; -5.86428 -i\,6.28319 \; $ 
& $\;\hphantom{-} 1.37392 - i\, 14.17999 \; $ \\
\hline
$\; {1_q^+} \,{2_g^+} \,{3_g^+} \,{4_g^+} \,{5_\qb^-} \,
   {6_{\lb_{\vphantom{A}}}^-} \,
  { 7_{l}^+} $
& $\;  -4.00000 \; $ 
& $\; -10.43958 - i\, 9.42478 \; $ 
& $\;\hphantom{-} 2.10030 - i\, 33.97042  \; $ \\
\hline
$\;{1_q^+} \,{2_g^+}\, {3_g^+}\, {4_g^-}\, {5_\qb^-}\, 
  {6_{\lb_{\vphantom{A}}}^-} \, {7_{l}^+} $
& $\;  -4.00000 \; $ 
& $\;  -10.43958 - i\, 9.42478 \; $ 
& $\;  -9.25860 - i\, 33.31407  \; $ \\
\hline
$\; {1_q^+} \,{2_g^-}\, {3_g^+}\, {4_g^+} \,{5_\qb^-}\, 
  {6_{\lb_{\vphantom{A}}}^-}\,   {7_{l}^+} $
& $\;  -4.00000   \; $ 
& $\; -10.43958 - i\, 9.42478  \; $ 
& $\; -5.51046   - i\, 33.55522  \; $ \\
\hline
$\; {1_q^+}\, {2_g^-} \,{3_g^+}\, {4_g^-}\, {5_\qb^-}\, 
 {6_{\lb_{\vphantom{A}}}^-}\, {7_{l}^+} $
& $\; -4.00000   \; $ 
& $\; -10.43958 - i\, 9.42478  \; $ 
& $\; -6.36853 -i \, 29.29380   \; $ \\
\hline
\hline
\end{tabular}
\end{center}
\end{table}

To test the implementation of vector-boson amplitudes in \BlackHat, we
first evaluated the $ q gg \qb \, \lb l $ one-loop primitive amplitudes
given in eqs.~(8.4)--(8.19) of ref.~\cite{Zqqgg}.  A sample diagram
for these primitive amplitudes is shown in \fig{Z45DiagramsFigure}(a).
The amplitudes; values from \BlackHat{}, at the phase-space point in
eq.~(9.1) of ref.~\cite{Genhel}, are presented on the first three
lines of table~\ref{VectorBosonTable}. We have normalized the values
by dividing by the corresponding tree amplitudes.  Similarly, at the
phase-space point in eq.~(9.3) of ref.~\cite{Genhel}, we compute the
primitive amplitudes with one additional gluon, shown in
\fig{Z45DiagramsFigure}(b), with the results displayed on the final
four lines of table~\ref{VectorBosonTable}.  (We remove the universal
prefactor, $\cg = \Gamma(1+\eps)\Gamma^2(1-\eps)/
[(4\pi)^{2-\eps}\Gamma(1-2\eps)]$, where $\eps = (4-D)/2$, from the
evaluation.)

\begin{figure}[t]
\centering
\includegraphics[width=14.5cm,clip]{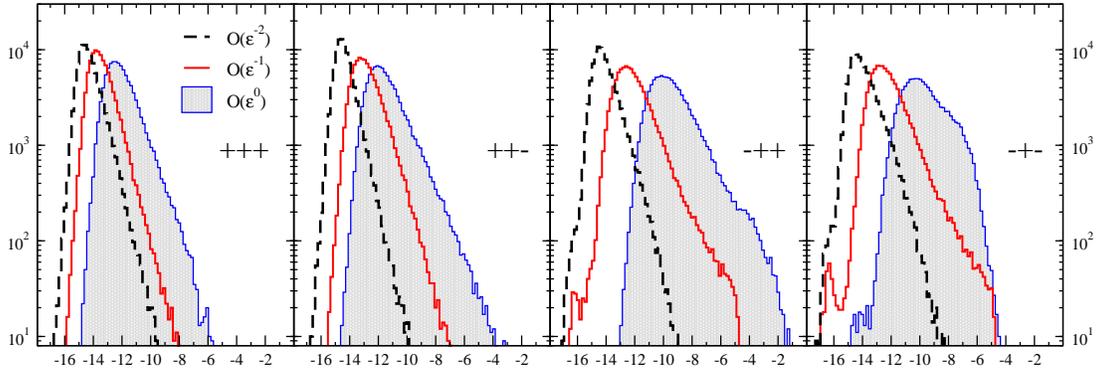}
\caption{The distribution of the logarithm of the relative error over 100,000
phase-space points, for four independent helicity choices for 
the  $q g g g \qb \, \lb l$ leading color amplitudes.  The helicity
labels on the plots indicate the helicities of the three gluons. The dashed
(black) curve in each histogram gives the relative error for the
$1/\eps^2$ part, the solid (red) curve gives the $1/\eps$ singularity,
and the shaded (blue) distribution gives the finite $\ep^0$ component
of the corresponding helicity amplitude. 
} 
\label{Z2q3gPlotsFigure}
\end{figure}

To assess the numerical stability of \BlackHat{}, we evaluated the
amplitudes in table~\ref{VectorBosonTable} at 100,000 phase-space
points drawn from a flat distribution, taking the quark and gluon
labeled by 1 and 2 as the incoming partons.  (We impose the following
cuts: $E_T > 0.01 \sqrt{s}$, $\eta<3$, and $\Delta_R >0.4$.)  The
histograms in \fig{Z2q3gPlotsFigure} show the results for the
seven-point amplitudes.  The horizontal axis gives the logarithmic
relative error,
$
\log_{10}({|A_n^{\rm num}-A_n^{\rm target}|} /{|A_n^{\rm target}|})\,,
$
for each of the $1/\e^2$, $1/\e$, and $\e^0$ parts of the one-loop
amplitude.  For the seven-point target---where no previous results are
available---we used quadruple-double ($\sim 64$ digits) precision
results evaluated in \BlackHat.  The vertical axis in these plots
shows the number of phase-space points in a bin that agree with the
target to a specified relative precision.  The vertical scale is
logarithmic, which enhances the visibility of the tail of the
distribution, illustrating the good numerical stability.


In summary, in this talk we presented the first, numerically stable
evaluation of the leading-color virtual matrix elements needed for the
NLO corrections to $p p \rightarrow V + 3$ jets, including vector
boson decay to a pair of leptons.  The subleading-color contributions
can be computed in a similar manner.  A key remaining task is to
interface \BlackHat{} with an automated
program~\cite{AutomatedSubtraction} for combining real and virtual
matrix elements, in order to produce NLO cross sections for
phenomenologically important multi-jet processes at the LHC.

We thank Academic Technology Services at UCLA for computer support.
This research was supported by the US Department of Energy under
contracts DE--FG03--91ER40662, DE--AC02--76SF00515 and DE--FC02--94ER40818. 
DAK's research is supported by the Agence
Nationale de la Recherce of France under grant
ANR--05--BLAN--0073--01.  The work of DM was supported by the Swiss
National Science Foundation (SNF) under contract PBZH2--117028.

\end{document}